\documentclass[manuscript]{aastex}
\usepackage{natbib}
\def\be{\begin{equation}}
\def\ee{\end{equation}}
\def\bdm{\begin{displaymath}}
\def\edm{\end{displaymath}}

\def\be{\begin{equation}}
\def\bdm{\begin{displaymath}}
\def\edm{\end{displaymath}}
\def\ebe{\end{displaymath}\begin{equation}}
\def\eba{\end{displaymath}\begin{displaymath}}

\def\a{\alpha }

\def\omr{\omega _R}

\def\wpe{\omega _{p,e}}
\def\Tp{\Theta _p}
\def\Te{\Theta _e}
\def\Tn{\Theta }
\def\upa{u_{a,\parallel}}
\def\2kua{\sqrt{2}ku_{a,\parallel}}
\def\bp{\beta _{\parallel }}
\def\w{w^2}
\def\a{\alpha }

\def\wp{weakly propagating }
\def\ct{cyctronic }
\def\mpm{\mu ^{1/2}}
\def\mi2{\mu ^{-1/2}}
\def\m32{\mu ^{3/2}}
\def\bc{\beta _c}
\shorttitle{Transverse temperature anisotropy instabilities in magnetized thermal plasmas}
\shortauthors{R. Schlickeiser \& T. Skoda}
\begin{document}

\title{Linear theory of weakly amplified, parallel propagating, transverse temperature anisotropy instabilities in magnetized thermal plasmas}
\author{R. Schlickeiser$^{1,2}$. T. Skoda$^1$}
\affil{1 Institut f\"ur Theoretische Physik, Lehrstuhl IV:
Weltraum- und Astrophysik, Ruhr-Universit\"at Bochum,
D-44780 Bochum, Germany\\
2 Research Department Plasmas with Complex Interactions, Ruhr-Universit\"at Bochum, D-44780 Bochum, Germany}
\email{rsch@tp4.rub.de, ts@tp4.rub.de}
\received{2010 February 1}
\begin{abstract}
A rigorous analytical study of the dispersion relations of weakly
amplified transverse fluctuations with wave vectors ($\vec{k}\parallel
\vec{B}$) parallel to the uniform background magnetic field $\vec{B}$
in an anisotropic bi-Maxwellian magnetized electron-proton plasma is
presented. A general analytical instability condition is derived that
holds for different values of the electron ($A_e$) and proton ($A_p$)
temperature anisotropies. We determine the conditions for which the
weakly amplified LH-handed polarized Alfven-proton-cyclotron and
RH-handed polarized Alfven-Whistler-electron-cyclotron branches can be
excited. For different regimes of the electron plasma frequeny phase
speed $w=\wpe /(kc)$ these branches reduce to the RH and LH polarized
Alfven waves, RH polarized high- and low-phase speed Whistler, RH
polarized proton and LH polarized electron cyclotron modes. Analytic
instability threshold conditions are derived in terms of the combined
temperature anisotropy $A=T_{\perp }/T_{\parallel }$, the parallel
plasma beta $\bp =8\pi n_ek_BT_{\parallel }/B^2$ and the electron
plasma frequency phase speed $w=\wpe /(kc)$ for each mode.

The results of our instability study are applied to the observed solar
wind magnetic turbulence at values of $90\le w\le 330$. According to
the existence conditions of the different instabilities, only the
left-handed and right-handed polarized Alfven wave instabilities can
operate here. Besides the electron-proton mass ratio $\mu =1836$, the
Alfvenic instability threshold conditions are controlled by the single
observed plasma parameter $w$. The Alfvenic instability diagram
explains well the main characteristic properties of the observed solar
wind fluctuations. Especially, the observed confinement limits at
small parallel plasma beta values are explained.

\end{abstract}
\keywords{plasmas -- instabilities -- turbulence -- magnetic fields  -- solar wind }
\section{Introduction}
Kinetic plasma relaxation and turbulence generation processes are
responsible for the observed properties of the solar wind plasma which
is the only cosmic collisionpoor plasma accessible to detailed in-situ
satellite observations \citep{Bale2009}. Although the detailed
plasma relaxation processes are not understood, the observed electron
and proton distribution functions are close to bi-Maxwellian
distributions with different temperatures along and perpendicular to
the ordered magnetic field direction. Ten years of Wind/SWE data
\citep{Kasper2002} have demonstrated that the proton and electron
temperature anisotropies $A=T_\perp/T_\parallel $ are bounded by
mirror and firehose instabilities \citep{Hellinger2006} at large
values of the parallel plasma beta $\beta _\parallel =8\pi
nk_BT_\parallel /B^2\ge 1$. In the parameter plane defined by the
temperature anisotropy $A=T_\perp/T_\parallel $ and the parallel
plasma beta $\beta _\parallel $, stable plasma configuration are only
possible within a rhomb-like configuration around $\beta _\parallel
\simeq 1$, whose limits are defined by the threshold conditions for
the mirror and firehose instabilities. If a plasma would start with
parameter values outside this rhomb-like configuration, it immediately
would generate fluctuations via the mirror and firehose instabilities,
which quickly relax the plasma distribution into the stable regime
within the rhomb-configuration. The plasma parameters of other dilute
cosmic plasmas including the interstellar and intracluster medium
\citep{sheko05} and accretion disks around compact, massive
objects \citep{sharma07} are similar to the solar wind plasma, so
that the temperature-anisitropy instabilities should also operate in
these systems.

The study of linear electromagnetic instabilities in a collisionless,
homogoneous, magnetized, electron-proton plasma with temperature
anisotropies has a long history; for reviews we refer the interested
reader to Ch. 7 of the monograph by \citet{g93}, \citet{c92} and
\citet{m06}. \citet{Hellinger2006} considered linear instability
calculations for the oblique mirror, firehose and proton cyclotron
instabilities, and represented the approximate threshold conditions
from the work of \citet{g94,gl01}, \citet{gl94},
\citet{s01} and \citet{p04} by analytic
relations of the form $A=1+a(\bp -\beta _0)^{-b}$ where $a$, $b$ and
$\beta _0$ are fitted parameter values.

In order to understand the confinement limits also at small values of
the parallel plasma beta $\beta _\parallel <1$, we analyze here
rigorously the full linear dispersion relation in a collisionless
homogenous plasma with anisotropic ($A\ne 1$) bi-Maxwellian particle
velocity distributions of electrons and protons for electromagnetic
fluctuations with wave vectors ($\vec{k}\times \vec{B}=0$) parallel to
the uniform background magnetic field $\vec{B}$. With a typical solar
wind temperature $T=10^5T_5$K, the thermal particle energy
$k_BT=8.6T_5$ eV is much less than the electron rest mass
$m_ec^2=5.11\cdot 10^5$ eV, so that the use of the nonrelativistic
linearized Vlasov/Maxwell equations is justified.  For weakly ($\gamma
\ll \omega _R$) amplified wave solutions we derive analytically
existence and instability conditions.  $\omega _R$ and $\gamma $ here
refer to real and imaginary part of the complex frquency $\omega
=\omega _R+\imath \gamma $.  Our analysis follows closely the recent
study \citep{s09} for equal-mass pair plasmas -- hereafter
referred to as paper S.  Our investigation is restricted to weakly
amplified ($\gamma \ll \omega _R$) solutions covering the left-handed
(LH) polarized Alfven-proton cyclotron branch and the right-handed
(RH) polarized Alfven-Whistler-electron cyclotron branch. The
corresponding analysis of \wp ($\omega _R\ll \gamma $) solutions,
including mirror, firehose, electron \ct and cold magnetized Weibel
fluctuations, is the subject of a subsequent paper. As we will
demonstrate below, for the case of parallel ($\vec{k}\times
\vec{B}=0$) wavevectors simple analytical threshold conditions for the
two branches can be deerived in terms of the combined temperature
anisotropy $A=T_{\perp }/T_{\parallel }$, the parallel plasma beta
$\bp =8\pi n_ek_BT_{\parallel }/B^2$, the electron-proton mass ratio
and the electron plasma frequency phase speed $w=\wpe /(kc)$.
\section{Dispersion relations}
\subsection{Basic equations}
For a nonzero background magnetic field strength the nonrelativistic
dispersion relations for right-handed (RH) and left-handed (LH)
polarized fluctuations with wave vectors $\vec{k}\times \vec{B}=0$ in
a thermal electron-proton plasma are \citep{g93}

\begin{eqnarray} 
0=D_{RH,LH}(k, \omega )&\!\!=\!\!& \omega^2-k^2c^2+\sum _{a=p,e}
\omega _{p,a} ^2\Bigg[\frac{\omega}{\sqrt{2}ku_{a,\parallel}}Z
\left(\frac{\omega \pm \Omega _a}{\sqrt{2}ku_{a,\parallel}}\right)\nonumber\\
&&\qquad+\frac{1}{2}\left(1-A_a\right) Z\sp{\prime}
\left(\frac{\omega \pm \Omega _a}{\sqrt{2}ku_{a,\parallel}}\right)\Bigg]=0, 
\label{a1}
\end{eqnarray}
where we sum over a proton (p)-electron (e) plasma. and where
$k=|k_{\parallel }|$.  $\omega _{p,e}$ denotes the electron plasma
frequency, $\upa =(k_BT_{a,\parallel }/m_a)^{1/2}$ is the parallel
thermal velocity of component $a$, $\Omega _a=e_aB/(m_ac)$ is the
non-relativistic gyrofrequency, and $A_a=T_{a, \perp }/T_{a, \parallel
}$ is the temperature anisotropy of component $a$, where the
directional subscripts refer to directions relative to the background
magnetic field.  The dispersion relations (\ref{a1}) allow for
different values of the proton and electron parallel temperatures and
temperature anisotropies.

$Z(x)$ and $Z\sp{\prime} (x)$ denote the plasma dispersion function
\citep{fc61} and its derivative

\be
Z(x)=\pi ^{-1/2}\int_{-\infty }^\infty dt\, {e^{-t^2}\over t-x}
\label{a3}
\ee
with the well-known properties 

\be
Z\sp{\prime} (x)=-2\left[1+xZ(x)\right],
\label{a4}
\ee
and 

\be 
Z(-x)=2\pi ^{1/2}\imath e^{-x^2}-Z(x),\;\;\; Z\sp{\prime} (-x)=4\pi ^{1/2}\imath xe^{-x^2}+Z\sp{\prime}(x)
\label{a41}
\ee
We will frequently use the asymptotic expansions 

\be
Z(x)\simeq \imath \pi ^{1/2}e^{-x^2}-\, 2x\left[1-{2x^2\over 3}\right], \; |x|\ll 1
\label{a5}
\ee
and 

\be
Z(x)\simeq \imath \sigma \pi ^{1/2}e^{-x^2}-\, {1\over x}\left[1+{1\over 2x^2}+{3\over 4x^4}\right], \; |x|\gg 1
\label{a6}
\ee
where $\sigma =0$ if $\Im (x)>0$, $\sigma =1$ if $\Im (x)=0$ and $\sigma =2$ if $\Im (x)<0$. 

Paper S has demonstrated that the analytical analysis is enormously
faoilitated if we work with space speed rather than frequencies. We
therefore introduce the complex phase speeds

\be
f={\omega \over kc}={\omega _R+\imath \gamma \over kc}=R+\imath S,\;\;\; 
R={\omr \over kc},\;\;\; S={\gamma \over kc},
\label{a7}
\ee
the plasma frequency phase speed 

\be
w={\wpe \over kc},
\label{a8}
\ee
and the absolute value of the electron gyrofrequeny phase speed 

\be
b={|\Omega _e|\over kc},
\label{a81}
\ee
where $|\Omega _e|=eB/m_ec$ is the absolute value of the electron
gyrofrequency. We also introduce the mass ratio

\be 
\mu =m_p/m_e=1836,
\label{a71}
\ee
and the dimensionless proton and electron temperatures 

\be
\Te \equiv \left({2k_BT_{e, \parallel}\over m_ec^2}\right)^{1/2},\;\; 
\Tp \equiv \left({2k_BT_{p, \parallel}\over m_pc^2}\right)^{1/2}
\label{a9}
\ee
Throughout this work, in the classification of \citet{sw89} we
discuss high density plasmas with $\wpe \gg |\Omega _e| $,
corresponding to $w\gg b$ which applies to nearly all astrophysical
plasmas. These plasmas are dense enough that the electron plasma
frequency is much larger than the electron gyrofrequency, but small
enough that elastic Coulomb collisions can be neglected.

The two dispersion relations (\ref{a1}) then read 

\bdm
0={D_{RH,LH}(k, f)\over k^2c^2}= \Lambda _{RH,LH}(k, f)=f^2-1+
\eba
{w^2\over \mu }\left[{f\over \Tp }Z\left({f\pm {b\over \mu }\over \Tp }\right)+
{1\over 2}\left(1-A_p\right)Z\sp{\prime} \left({f\pm {b\over \mu }\over \Tp }\right)\right]
\ebe
+w^2\left[{f\over \Te }Z\left({f\mp b\over \Te }\right)+{1\over 2}\left(1-A_e\right)Z\sp{\prime} \left({f\mp b\over \Te}\right)\right]
\label{a10}
\ee
We notice the symmetry $\Lambda (-k_{\parallel},f)=\Lambda
(k_{\parallel },f)$ of both dispersion relations allowing to consider
only positive values of the wavenumber $k>0$. In the following we will
simplify the analysis by considering only equal parallel temperature
plasmas ($T_{e,\parallel }=T_{p,\parallel }$) so that $\Te =\Tn $ and
$\Tp =\Tn /\mu ^{1/2}.$

The dispersion relations (\ref{a10}) can be separated into real and
imaginary parts $\Lambda (R, S)=\Re \Lambda (R, S)\, +\imath \Im
\Lambda (R, S)=0$, implying the two conditions \be \Re \Lambda (R,
S)=0,\, \;\; \Im \Lambda (R, S)=0
\label{aa2}
\ee
In terms of the complex phase speed $f=R+\imath S$ the real and
imaginary parts of the two dispersion relations (\ref{a10}) read

\bdm
0=\Re \Lambda _{RH,LH}(R,S)=R^2-S^2-1+
\eba
\w \Bigl[{R\over \Tn \mu ^{1/2}}\Re Z\left({\mu ^{1/2}\over \Tn}\left[R+\imath S\pm {b\over \mu }\right]\right)
+{1-A_p\over 2\mu }\Re Z\sp{\prime} \left({\mu ^{1/2}\over \Tn}\left[R+\imath S\pm {b\over \mu }\right]\right)
\eba
+{R\over \Tn }\Re Z\left({R+\imath S\mp b\over \Tn}\right)
+{1-A_e\over 2}\Re Z\sp{\prime} \left({R+\imath S\mp b\over \Tn }\right)\Bigr]
\ebe
-{\w S\over \Tn }\left[{1\over \mu ^{1/2}}\Im Z\left({\mu ^{1/2}\over \Tn}\left[R+\imath S\pm {b\over \mu }\right]\right)
+\Im Z\left({R+\imath S\mp b\over \Tn}\right)\right]
\label{a12}
\ee
and

\begin{eqnarray}
\!\!\!\!\!\!0=\Im \Lambda _{RH,LH}(R, S)&\!\!\!=\!\!\!&2RS+
{\w \over 2}\Bigg[{1-A_p\over \mu }\Im Z\sp{\prime} \left({\mu ^{1/2}\over \Tn}\left[R+\imath S\pm {b\over \mu }\right]\right)\nonumber\\
&&\quad+(1-A_e)\Im Z\sp{\prime} \left({R+\imath S\mp b\over \Tn}\right)\Bigg]\nonumber\\
&&\quad+{\w S\over \Tn }\left[{1\over \mu ^{1/2}}\Re Z\left({\mu ^{1/2}\over \Tn}\left[R+\imath S\pm {b\over \mu }\right]\right)+
\Re  Z\left({R+\imath S\mp b\over \Tn}\right)\right]\nonumber\\
&&\quad+{\w R\over \Tn }\left[{1\over \mu ^{1/2}}\Im Z\left({\mu ^{1/2}\over \Tn}\left[R+\imath S\pm {b\over \mu }\right]\right)+
\Im Z\left({R+\imath S\mp b\over \Tn}\right)\right]
\label{a13}
\end{eqnarray}
Here we consider solutions of the dispersion relations in the weak
damping/amplification limit $|S|\ll R$. As shown in paper S in this
limit the real part of the dispersion relation satisfies

\be
\Re \Lambda (R, S=0)=0,
\label{aa3}
\ee
whereas the corresponding imaginary part is given by 

\be
S=-{\Im \Lambda (R, S=0)\over {\partial \Re \Lambda (R, S=0)\over \partial R}}
\label{aa4}
\ee
As before we introduce the parallel plasma beta 

\be
\bp ={\Tn ^2\w \over b^2}={8\pi n_ek_BT_{\parallel }\over B^2},
\label{a14}
\ee
which expresses the magnetic field strength as 

\be
b={\Tn w\over \bp ^{1/2}}
\label{a15}
\ee
in terms of the electron plasma frequency ($w$), the parallel
temperature ($\Tn $) and the parallel plasma beta $\bp $. Our
restriction to high-density plasmas with $\wpe >|\Omega _e|$ or $w>b$
then requires to have parallel plasma betas $\bp >\Tn ^2$.

\section{Weakly damped and amplified solutions}
For weakly damped or amplified fluctuations Eqs. (\ref{a12}) -- (\ref{a13}) read

\bdm
0=\Re \Lambda _{RH,LH}(R, S=0)=R^2-1
\eba
+\w \Bigl[{R\over \Tn }\Re Z\left({R\mp b\over \Tn}\right)
+{1-A_e\over 2}\Re Z\sp{\prime} \left({R\mp b\over \Tn }\right)
\ebe
+{R\over \Tn \mu ^{1/2}}\Re Z\left({\mu ^{1/2}\over \Tn}\left[R\pm {b\over \mu }\right]\right)
+{1-A_p\over 2\mu }\Re Z\sp{\prime} \left({\mu ^{1/2}\over \Tn}\left[R\pm {b\over \mu }\right]\right)\Bigr]
\label{b1}
\ee
and

\bdm
\Im \Lambda _{RH,LH}(R, S=0)=
{\w \over 2}\left[{1-A_p\over \mu }\Im Z\sp{\prime} \left({\mu ^{1/2}\over \Tn}\left[R\pm {b\over \mu }\right]\right)+
(1-A_e)\Im Z\sp{\prime} \left({R\mp b\over \Tn}\right)\right]
\edm
\be
+{\w R\over \Tn }\left[{1\over \mu ^{1/2}}\Im Z\left({\mu ^{1/2}\over \Tn}\left[R\pm {b\over \mu }\right]\right)+
\Im Z\left({R\mp b\over \Tn}\right)\right],
\label{b2}
\ee
which have to be investigated for positive values of $R\ge 0$. 

As noted before, these dispersion relations can be further reduced
with the asymptotic expansions (\ref{a5}) and (\ref{a6}), depending on
the absolute values of the arguments

\be
P_{\pm }(R)={\mu ^{1/2}\over \Tn}\left[|R\pm {b\over \mu }|\right],\;\; E_{\pm }(R)={|R\pm b|\over \Tn } 
\label{b3}
\ee
of the plasma dispersion function and its derivative being small or
large compared to unity. Scaling $R=bx$, corresponding to $x=\omega
_R/|\Omega _e|$, the absolute values of the arguments (\ref{b3}) in
terms of the parallel plasma beta read

\be
P_{\pm }(x)={w\mu ^{1/2}\over \bp^{1/2}}\left[|x\pm {1\over \mu }|\right],\;\; E_{\pm }(x)={w\over \bp^{1/2}}|x\pm 1|,
\label{b4}
\ee
which are shown in Fig. 1 as a function of the normalized real frequency $x$. 

\begin{figure}[t]
\begin{center}
\includegraphics[width=160mm]{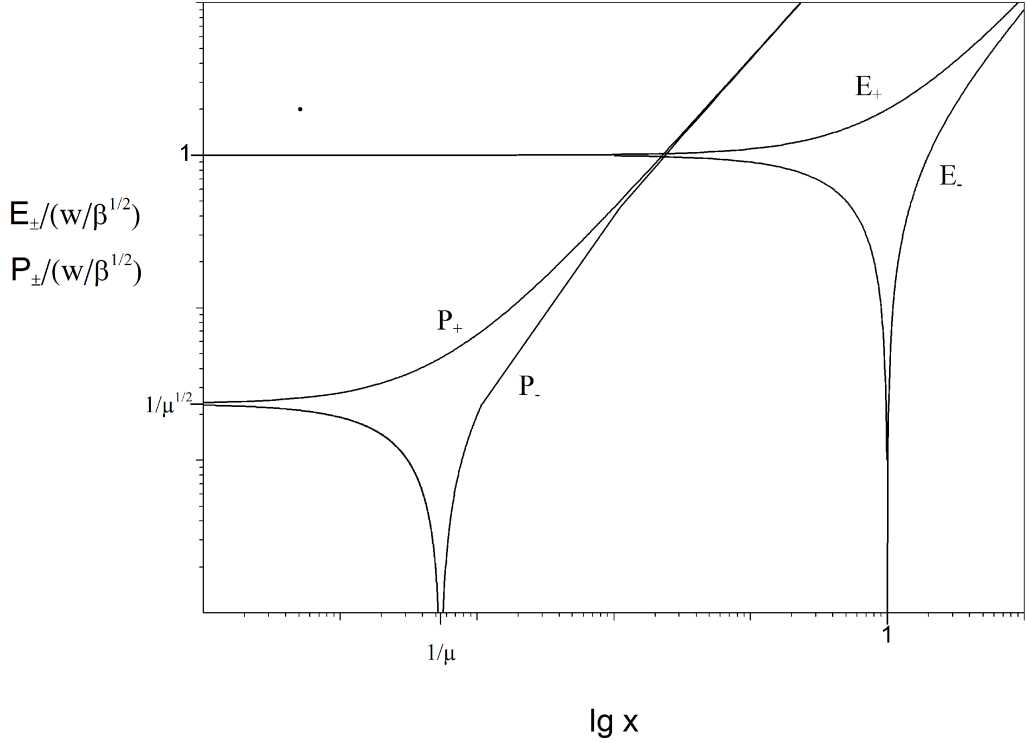}
\end{center}
\caption{Absolute values of the arguments $P_{\pm }(x)$ and $E_{\pm
  }(x)$ of the plasma dispersion function ans its derivative in the
  weak damping/amplification limit as a function of $x=R/b$.}
\end{figure}
For parallel plasma beta values $\bp \ll (w^2/\mu )$ we note that
$P_+(x)\gg 1$, $E_+(x)\gg 1$ for all values of $x$, whereas $P_-(x)\gg
1$ for $x$ outside the small interval

\be
x\notin \left[{1\over \mu }-{\bp^{1/2}\over w\mu^{1/2}}, {1\over \mu }+{\bp^{1/2}\over w\mu^{1/2}}\right]
\label{b5}
\ee
around the proton cyclotron frequency, and $E_-(x)\gg 1$ for $x$
outside the small interval

\be
x\notin \left[1-{\bp^{1/2}\over w}, 1+{\bp^{1/2}\over w}\right]
\label{b6}
\ee
around the electron cyclotron frequency. 

In the following we limit our analysis to such values of the parallel
plasma beta $\bp \ll (w^2/\mu )$, so that the asymptotic expansion
(\ref{a6}) of the plasma dispersion function can be used except near
the indicated proton and electron cyclotron frequencies.

Irrespective of the values of $P_{\pm }$ and $E_{\pm }$ we note that
both asymptotic expansions (\ref{a5}) and (\ref{a6}) yield the same
imaginary part of the dispersion relation

\bdm
0=\Im \Lambda _{RH,LH}(R, S=0)=\pi ^{1/2}\w {b\over \Tn }
\Bigl({1\over \mu ^{1/2}}\left[A_p\left[{R\over b}\pm {1\over \mu }\right]\mp {1\over \mu }\right]
e^{-{{\mu \over \Tn ^2}(R\pm {b\over \mu })^2}}
\edm
\be
+\left[A_e\left[{R\over b}\mp 1\right]\pm 1\right]e^{-{(R-b\over \Tn })^2}\Bigr),
\label{b9}
\ee
\section{Alfven, Whistler, cyclotron waves and electromagnetic light} 
For parallel plasma beta values $\bp \ll (w^2/\mu )$ the asymptotic
expansion (\ref{a6}) yields

\be
\Re Z\left({R\pm b\over \Tn }\right)\simeq -{\Tn \over R\pm b}\left[1+{\Tn ^2\over 2(R\pm b)^2}\right],
\label{b10}
\ee

\be
\Re Z\sp{\prime} \left({R\pm b\over \Tn}\right)\simeq 
+{\Tn ^2\over (R\pm b)^2}\left[1+{3\Tn ^2\over 2(R\pm b)^2}\right],
\label{b11}
\ee

\be
\Re Z\left({\mu ^{1/2}\over \Tn}\left[R\pm {b\over \mu }\right]\right)\simeq 
-{\Tn \over \mu ^{1/2}(R\pm {b\over \mu })}\left[1+{\Tn ^2\over 2\mu (R\pm {b\over \mu })^2}\right]
\label{b12}
\ee
and

\be
\Re Z\sp{\prime} \left({\mu ^{1/2}\over \Tn}\left[R\pm {b\over \mu }\right]\right)\simeq 
+{\Tn ^2\over \mu (R\pm {b\over \mu })^2}\left[1+{3\Tn ^2\over 2\mu (R\pm {b\over \mu })^2}\right]
\label{b13}
\ee
We then obtain for the real part of the dispersion relations
(\ref{b1}) to lowest order in $\Tn ^2\ll 1 $

\bdm
0=\Re \Lambda _{RH,LH}(R, S=0)=R^2-1-\w R\left[{1\over R\mp b}+{1\over \mu (R\pm {b\over \mu })}\right]
\edm
\be
+{\Tn ^2\w \over 2}\left[{1-A_e\over (R\mp b)^2}-{R\over (R\mp b)^3}\right]
+{\Tn ^2\w \over 2\mu ^2}\left[{1-A_p\over (R\pm {b\over \mu })^2}-{R\over (R\pm {b\over \mu })^3}\right]
\label{b14}
\ee
With the scaling $R=bx$ the dispersion relation (\ref{b14}) reads

\bdm
0=\Re \Lambda _{RH,LH}(x, S=0)=\left[b^2+{(1+\mu )\w \over (1\pm \mu x)(1\mp x)}\right]x^2-1
\ebe
\pm {\bp \over 2}\left({A_ex\pm (1-A_e)\over (1\mp x)^3}-{A_p\mu x\mp (1-A_p)\over (1\pm \mu x)^3}\right)
\label{b15}
\ee
where we introduce the parallel plasma beta (\ref{a14}).

In different limits the solutions of Eqs. (\ref{b14}) and (\ref{b15})
describe Alfven waves, Whistler waves, cyclotron waves and
electromagnetic light. We consider each case in the next sections.
\subsection{Alfven waves at phase speeds $R\ll b/\mu $ for $\Tn ^2\ll \bp <\w /\mu $}
For phase speeds $R\ll b/\mu $ the dispersion relation (\ref{b14})
simplifies to

\begin{eqnarray}
0=\Re \Lambda _{RH,LH}(R, S=0)&\!\!\!=\!\!\!&R^2\left(1+{2\w (1+\mu )\over b^2}
\right)-1\nonumber \\ 
&& \quad +{\bp \over 2}\left[2-A_e-A_p\pm {2R\over
    b}(3-A_e)\mp {2\mu R\over b}(3-A_p)\right]\nonumber \\ 
&&\simeq R^2\left(1+{c^2\over V_A^2}\right)-\left[1+\left(A-1\right)\bp \right],
\label{c1}
\end{eqnarray}
where we introduce the Alfven speed 

\be
{(1+\mu )\w \over b^2}={4\pi n_e(m_e+m_p)c^2\over B^2}={c^2\over V_A^2},
\label{c2}
\ee
the parallel plasma beta (\ref{a14}) and the combined plasma
temperature anisotropy

\be
A={A_p+A_e\over 2}
\label{c3}
\ee
For high-density plasmas $V_A\ll c$, the dispersion relation
(\ref{c1}) yields the LH and RH polarized Alfven modes with the same

\be
R\simeq {b\over \sqrt{1+\mu }w}\sqrt{1+(A-1)\bp }=\Tn \sqrt{{1+(A-1)\bp \over (1+\mu )\bp }}
={V_A\over c}\sqrt{1+(A-1)\bp },
\label{c4}
\ee
which either can propagate forward and backward. Note that the
condition $R\ll b/\mu $ requires $(\w /\mu )\gg 1$.

Moreover, the four Alfven modes only exist for temperature
anisotropies such that $1+(A-1)\bp \ge 0$ corresponding to

\be
A>\left(1-{1\over \bp }\right)
\label{c5}
\ee
which includes the isotropic ($A_p=A_e=A=1$) plasma temperature
case. For small plasma betas ($\bp \le 1$) the condition (\ref{c5}) is
always fulfilled whereas for large plasma betas ($\bp >1$) the
combined anisotropy $A$ has to be larger than $1-\bp ^{-1}$.

Eq. (\ref{c1}) also provides 

\be
{\partial \Re \Lambda _{RH,LH}(R, S=0)\over \partial R}
\simeq 2{c^2\over V_A^2}R,
\label{c6}
\ee
so that for all four modes according to Eqs. (\ref{aa4}) and
(\ref{b9}) the growth/damping rate is

\bdm
S_{RH,LH}=-{V_A^2\over 2c^2R}\Im \Lambda _{RH,LH}(R, S=0)=
\edm
\be
-{\pi ^{1/2}b^3\over 2(1+\mu )\Tn R}
\Bigl({1\over \mu ^{1/2}}\left[A_p\left[{R\over b}\pm {1\over \mu }\right]\mp {1\over \mu }\right]
e^{-{{\mu \over \Tn ^2}(R\pm {b\over \mu })^2}}
+\left[A_e\left[{R\over b}\mp 1\right]\pm 1\right]e^{-{(R\mp b\over \Tn })^2}\Bigr)
\label{c7}
\ee
For isotropic ($A_p=A_e=1$) plasma temperatures all four Alfven modes
are damped in agreement with the general theorem of \citet{br90} on the
electromagnetic stability of isotropic plasma populations. The Alfven
damping rate in the isotropic case is given by

\be
S_{RH,LH}(A_p=A_e=1)=-{\pi ^{1/2}b^2\over 2(1+\mu )\Tn }
\left[{1\over \mu ^{1/2}}e^{-{{\mu \over \Tn ^2}(R\pm {b\over \mu })^2}}+
e^{-{(R\mp b\over \Tn })^2}\right]
\label{c8}
\ee
In order to drive the Alfven modes unstable, the growth rate
(\ref{c7}) has to be positive requiring that

\bdm
I_{RH,LH}(A_p,A_e,\bp )\equiv \pm \left({A_p-1\over \mu ^{3/2}}e^{-{\w \over \mu \bp }(1\pm \mu x)^2}
-(A_e-1)e^{-{\w \over \bp }(1\mp x)^2}\right)
\ebe
+x\Bigl[{A_p\over \mu ^{1/2}}e^{-{\w \over \mu \bp }(1\pm \mu x)^2}
+A_ee^{-{\w \over \bp }(1\mp x)^2}\Bigr]<0
\label{c9}
\ee
This instability condition is analyzed further in the next section.
\subsection{LH-polarized Alfven-proton-cyclotron and RH-polarized Alfven-Whistler-electron cyclotron branches 
for small parallel plasma beta $\bp \ll 1$}
Because we are primarily interested in low plasma beta plasmas we
consider for completeness the solutions of the full dispersion
relation (\ref{b15}) for subluminal solutions $R\ll 1$ and small
parallel plasma beta $\bp \ll 1$:

\be
0=\Re \Lambda _{RH,LH}(x, S=0)\simeq {(1+\mu )\w x^2\over 1\pm (\mu -1)x-\mu x^2}-1
\label{d1}
\ee
yielding the quadratic equation 

\be
\left[(1+\mu )\w +\mu \right]x^2\mp (\mu-1)x=1
\label{d2}
\ee
Because $\mu =1836\gg 1$ this equation is well approximated by 

\be
(1+\w )x^2\mp x={1\over \mu }
\label{d3}
\ee
with the solutions 

\be
x_{RH,LH}={1\over 2(1+\w )}\left[\sqrt{1+{4(1+\w )\over \mu }}\pm 1\right],
\label{d4}
\ee
covering the well-known \citep[e.g.][]{sw89} LH-polarized
Alfven-proton-cyclotron and RH-polarized Alfven-Whistler-electron
cyclotron branches for parallel propagation.

For $w\ll \sqrt{(\mu /4)-1}=21.4$, corresponding to large wavenumbers
$k\gg \wpe /21.4c$, we obtain

\be
x_{RH}\simeq {1\over \mu }+{1\over 1+\w }
\label{d5}
\ee
and 

\be
x_{LH}\simeq {1\over \mu }[1-{1+\w \over \mu }]
\label{d6}
\ee
The solution (\ref{d6}), subject to the constraint (\ref{b5}),
represents the left-handed polarized proton-cyclotron waves.

For values of $w\ll 21.4$ the solution (\ref{d5}) reduces to 

\be
x_{RH}\simeq {1\over 1+w^2},
\label{d7}
\ee
which for values of $w\ll 1$ represents the right-handed polarized
electron cyclotron waves subject to the constraint (\ref{b6}) with the
dispersion relation

\be
x_{RH}\simeq 1-w^2.
\label{d71}
\ee
For intermediate values $1\ll w\ll 21.4$ the solution (\ref{d7})
reduces to

\be
x_{RH}\simeq {1\over \w },
\label{d8}
\ee
representing the right-handed polarized Whistler waves ($R\simeq b/\w
$).

For $w\gg \sqrt{(\mu /4)-1}=21.4$, corresponding to small wavenumbers
$k\ll \wpe /21.4c$, we obtain for the solution (\ref{d4})

\be
x_{RH,LH}\simeq {1\over \sqrt{\mu (1+\w )}}\simeq {1\over \mu ^{1/2}w},
\label{d9}
\ee
which agrees with the Alfven wave solutions (\ref{c4}) in the limit
$\bp \ll 1$.

Eq. (\ref{d1}) for small plasma beta $\bp \ll 1$ also provides 

\be
{\partial \Re \Lambda _{RH,LH}(R, S=0)\over \partial R}\simeq 
{(1+\mu )\w x\over b}{2\pm (\mu -1)x\over [1\pm (\mu -1)x-\mu x^2]^2}
={2\pm (\mu -1)x\over b(1+\mu )\w x^3},
\label{d10}
\ee
We note that for all solutions (\ref{d5}) -- (\ref{d9}),
Eq. (\ref{d10}) is positive. According to Eqs. (\ref{aa4}) and
(\ref{b9}) the growth/damping rate then is

\bdm
S_{RH,LH}=-{b(1+\mu )\w x^3\over 2\pm (\mu -1)x}\Im \Lambda _{RH,LH}(R, S=0)=
\edm
\be
-{\pi ^{1/2}(1+\mu )b^2w^4x^3\over \Tn [2\pm (\mu -1)x]}
\Bigl({1\over \mu ^{1/2}}\left[A_p\left[x\pm {1\over \mu }\right]\mp {1\over \mu }\right]
e^{-{{\mu b^2\over \Tn ^2}(x\pm {1\over \mu })^2}}
+\left[A_e(x\mp 1)\pm 1\right]e^{-{b^2(x\mp 1)^2\over \Tn ^2}}\Bigr),
\label{d11}
\ee
which generalizes the Alfvenic growth/damping rate (\ref{c7}) to
non-Alfvenic modes. Apart from the different notation in phase speeds,
the rate (\ref{d11}) agrees with eq. (7.1.8) of \citet{g93}. For
isotropic ($A_p=A_e=1$) plasma temperatures all modes of the RH and LH
branches, including the cyclotron and Whistler modes, are damped in
agreement with Brinca's general theorem. We notice that, in order to
drive the cyclotron and Whistler modes unstable, the growth rate
(\ref{d11}) has to be positive, requiring again the earlier derived
instability condition (\ref{c9}).

\subsection{Electromagnetic light at large frequencies $R\gg b+\Tn $}
For large frequencies $R\gg b+\Tn $ the dispersion relation
(\ref{b14}) reduces to

\bdm
0=\Re \Lambda _{RH,LH}(R, S=0)\simeq R^2-1-(1+{1\over \mu })w^2 \mp w^2{b\over R}
-{\w b^2\over R^2}\left[1+{1\over \mu }+{\Tn ^2\over 2b^2}\left(A_e+{A_p\over 4\mu ^2}\right)\right]
\ebe
\simeq R^2-1-(1+{1\over \mu })w^2 
\label{d12}
\ee
and 

\be
0=\Im \Lambda _{RH,LH}(R, S=0)\simeq \pi ^{1/2}\w {R\over \Tn }\left({A_p\over \mu ^{1/2}}e^{-\mu R^2/\Tn ^2}+A_ee^{-R^2/\Tn ^2}\right),
\label{d13}
\ee
so that for both polarisations to lowest order in $(b/R)^2$ we obtain 
the dispersion relation of electromagnetic light 

\be
R^2\simeq 1+(1+{1\over \mu })\w 
\label{d14}
\ee
with the same damping rate 

\be
S=-{\pi ^{1/2}\w \over 2\Tn }\left({A_p\over \mu ^{1/2}}e^{-\mu R^2/\Tn ^2}+A_ee^{-R^2/\Tn ^2}\right)
\label{d15}
\ee
\section{Instability conditions}
In order to drive RH-handed polarized Alfven-proton cylotron and the
LH-handed polarized Alfven-Whistler-electron cyclotron branches
unstable, the condition (\ref{c9}) has to be fulfilled which, after
multiplication with $\mu ^{3/4}$, reads

\bdm
\pm \left({A_p-1\over \mu ^{3/4}}e^{-X_p^2}-(A_e-1)\mu ^{3/4}e^{-X_e^2}\right)+
x\mu ^{1/2}\left[{A_p\over \mu ^{1/4}}e^{-X_p^2}+A_e\mu ^{1/4}e^{-X_e^2}\right]=
\eba
\pm \left(e^{-{3\over 4}\ln \mu +\ln (A_p-1)-X_p^2}-e^{{3\over 4}\ln \mu +\ln (A_e-1)-X_e^2}\right)
\ebe
+x\mu ^{1/2}\left(e^{-{1\over 4}\ln \mu +\ln A_p-X_p^2}+e^{{1\over 4}\ln \mu +\ln A_e-X_e^2}\right)<0
\label{e1}
\ee
with 

\be
X_p={w \over (\mu \bp )^{1/2}}(1\pm \mu x),\;\; X_e={w \over \bp ^{1/2}}(1\mp x).
\label{e2}
\ee
Both brackets in Eq. (\ref{e1}) can be further reduced using the
identities

\bdm
e^{a_1+a_2}+e^{a_1-a_2}=2e^{a_1}\cosh (a_2),\;\; e^{a_1+a_2}-e^{a_1-a_2}=2e^{a_1}\sinh (a_2),
\edm
\be
e^{c_1+c_2}+e^{c_1-c_2}=2e^{c_1}\cosh (c_2),\;\; e^{c_1+c_2}-e^{c_1-a_c}=2e^{c_1}\sinh (c_2)
\label{e3}
\ee
From the first bracket we identify 

\be
a_1=\ln \left[(A_p-1)(A_e-1)\right]^{1/2}-{X_p^2+X_e^2\over 2},\,\; 
a_2=-{3\over 4}\ln \mu -\ln \left[{A_e-1\over A_p-1}\right]^{1/2}-W_{\pm }
\label{e4}
\ee
where 

\be
W_{\pm }(x)={X_p^2-X_e^2\over 2}={\w \over 2\bp }\left[\pm 4x+(\mu -1)(x^2-{1\over \mu })\right],
\label{e5}
\ee
From the second bracket we determine 

\be
c_1=\ln \left[A_pA_e\right]^{1/2}-{X_p^2+X_e^2\over 2},\,\; 
c_2=-{1\over 4}\ln \mu -\ln \left[{A_e\over A_p}\right]^{1/2}-W_{\pm }
\label{e6}
\ee
With $\sinh (-a_2)=-\sinh (a_2)$ and $\cosh (-c_2)=\cosh (c_2)$
condition (\ref{e1}) becomes

\bdm
\pm (A_p-1)^{1/2}(A_e-1)^{1/2}\sinh \left[{3\over 4}\ln \mu +\ln \left[{A_e-1\over A_p-1}\right]^{1/2}+W_{\pm }\right]
\eba
-A_p^{1/2}A_e^{1/2}x\mu ^{1/2}\cosh \left[{1\over 4}\ln \mu +\ln \left[{A_e\over A_p}\right]^{1/2}+W_{\pm }\right]
\edm
\bdm
=\pm \left(\mu ^{3/4}(A_e-1)-\mu ^{-3/4}(A_p-1)\right)\cosh W_{\pm }\pm \left(\mu ^{3/4}(A_e-1)+\mu ^{-3/4}(A_p-1)\right)\sinh W_{\pm }
\edm
\be
-x\mu ^{1/2}\left(\mu ^{1/4}A_e+\mu ^{-1/4}A_p\right)\cosh W_{\pm }-x\mu ^{1/2}\left(\mu ^{1/4}A_e-\mu ^{-1/4}A_p\right)\sinh W_{\pm }>0,
\label{e7}
\ee
which yields the general instability condition for the two branches in
the form

\bdm
\pm {(A_e-A_p)[(\mu ^{3/2}+1)+(\mu ^{3/2}-1)\tanh (W_{\pm }]+(A_e+A_p-2)[(\mu ^{3/2}-1)+(\mu ^{3/2}+1)\tanh (W_{\pm }]\over 
x\mu \left((A_e+A_p)[\mpm +1+(\mpm -1)\tanh (W_{\pm })]+(A_e-A_p)[\mpm -1+(\mpm +1)\tanh (W_{\pm })]\right)}
\ebe
>1
\label{e8}
\ee
Inserting the general dispersion relation for small plasma betas
(\ref{d4}) for $x$ and $W_{\pm }(x)$ then provides the general
instability conditions for the RH-handed polarized Alfven-proton
cylotron and the LH-handed polarized Alfven-Whistler-electron
cyclotron branches, which to the best of our knowledge has not been
derived before.

For equal electron and proton temperature anisotropies ($A_p=A_e=A_0$)
the condition (\ref{e8}) reduces to

\bdm
\pm \left(1-{1\over A_0}\right)\left[(\mu ^{3/2}-1)+(\mu ^{3/2}+1)\tanh (W_{\pm }\right]
\ebe
>\mu x\left[\mpm +1+(\mpm -1)\tanh (W_{\pm })\right].
\label{e9}
\ee
Note that for pair plasmas ($\mu =1$) the condition (\ref{e9}) reduces
to Eq. (S-66) in paper S.

Instead of working with the general dispersion relations (\ref{d4})
used in the general instability condition (\ref{e8}), we will discuss
different ranges of the plasma frequency phase speed $w$, where the
dispersion relations (\ref{d4}) reduce to simpler limits, as
demonstrated in Sect. 4.2. For $w\gg 21.4$ we can use the Alfvenic
relation (\ref{d9}), whereas for small values of $w<21.4$ we can use
the cyclotron and Whistler relations (\ref{d6}), (\ref{d7}),
(\ref{d71}) and (\ref{d8}), respectively.

Before investigating the individual modes, we inspect the instability
condition (\ref{e9}). The instability conditions holds for plasma
betas $\bp \ll \w /\mu $, implying $\w /2\bp \gg \mu
/2=918$. Consequently, the argument of the $\tanh $-function

\be
W_{\pm }={\w \over 2\bp }f_{\pm }(x),\;\; f_{\pm }(x)=\pm 4x+(\mu -1)(x^2-{1\over \mu })
\label{e11}
\ee
is much larger than unity for $f_{\pm }(x)\gg 2/\mu $.  

The function $f_+(x)$ increases monotonically from its smallest
negative value $f_+(0)=-(1-\mu ^{-1})=-0.9995$ and becomes zero at

\be
x_{0+}={\mu +1\over (\mu -1)\mu ^{1/2}}\left[1-{2\mu ^{1/2}\over \mu +1}\right]=0.95\mu ^{-1/2}\simeq \mu ^{-1/2}
\label{e12}
\ee
Likewise, the function $f_-(x)$ attains its negative minimum value
$-1.0016$ at $x_E=2/(\mu -1)=0.0011$ and becomes zero at

\be
x_{0-}={\mu +1\over (\mu -1)\mu ^{1/2}}\left[1+{2\mu ^{1/2}\over \mu +1}\right]=1.05\mu ^{-1/2}\simeq \mu ^{-1/2}
\label{e13}
\ee
Hence only for values of $x=R/b$ inside a small interval around $\mu
^{-1/2}$, which for right-handed polarized fluctuations lies in the
Whistler phase speed range, the functions $f_{\pm }(x)$ are smaller
than $2/\mu $. Both functions are well approximated by

\be
f_{\pm }(x)\simeq f_0(x)=\mu x^2-1,
\label{e14}
\ee
which is negative in the range of Alfven and proton cyclotron waves
and positive in the range of electron cyclotron waves, respectively.
\subsection{Left-handed polarized Alfven waves and proton cyclotron waves}
In the range of Alfven and proton cyclotron waves $x\le \mu ^{-1}$ the
function $f_-(x)\simeq -1$ is negative and practically constant, so
that the instability condition (\ref{e9}) becomes

\bdm
\left(1-{1\over A_0}\right)\left[(\m32 +1)\tanh \left[{\w \over 2\bp }\right]-(\m32 -1)\right]
\ebe
>x\mu \left[(\mpm +1)-(\mpm -1)\tanh \left[{\w \over 2\bp }\right]\right]
\label{e15}
\ee
The $\tanh$-function is well approximated by 

\be
\tanh (t)\simeq {t\over 1+t},
\label{e16}
\ee
so that with  $2\bp /\w \ll 2/\mu =1/918\ll 1$

\be
\tanh \left[{\w \over 2\bp }\right]\simeq {1\over 1+{2\bp \over \w }}\simeq 1-{2\bp \over \w },
\label{e17}
\ee
we find for condition (\ref{e15}) 

\be
\left(1-{1\over A_0}\right)\left[1-(1+\m32 ){\bp \over \w }\right]>\mu x\left[1+(\mpm -1){\bp \over \w }\right].
\label{e18}
\ee
Because $\m32 \gg 1$ and $\mpm \gg 1$ condition (\ref{e18}) is well
approximated by

\be
(1-{1\over A_0})[1-\eta ]>x[\mu +\eta ],
\label{e19}
\ee
where 

\be
\eta ={\bp (\m32 +1)\over \w }=\bp /\beta _c
\label{e20}
\ee
denotes a normalized parallel plasma beta value in terms of the
critical plasma beta

\be
\beta _c={\w \over (1+\m32 )}
\label{e21}
\ee
The restriction $\bp \ll \w /\mu $ requires $\eta \ll \mpm =43$.
Obviously, we have to consider the two cases $\eta <1$ and $1<\eta
<43$, corresponding to $\bp <\beta _c$ and $\beta_c<\bp <43\beta _c$,
respectively.

For $\bp <\bc $ the instability condition (\ref{e19}) becomes

\be
1-{1\over A_0}>x{\mu +\eta \over 1-\eta },
\label{e22}
\ee
or 

\be
A_0(\bp <\bc )>{1-\eta \over (1-\mu x)-(1+x)\eta }
\label{e23}
\ee
Likewise, for $\bc <\bp <43\bc $ we obtain 

\be
({1\over A_0}-1)>x{\mu +\eta \over \eta -1},
\label{e24}
\ee
corresponding to 

\be
A_0(\bp >\bc )<{\eta -1\over (1+x)\eta -(1-\mu x)}
\label{e25}
\ee
For plasma beta values equal to the critical value ($\eta =1$), the
instability condition (\ref{e19}) cannot be fulfilled.  The two
instability conditions (\ref{e23}) and (\ref{e25}) for the LH
polarized Alfven and proton cyclotron waves are illustrated in Fig. 2
together with the instability conditions for the RH polarized Alfven
and low phase speed Whistler waves, which we derive in the next
subsection.

\begin{figure}[t]
\begin{center}
\includegraphics[width=160mm]{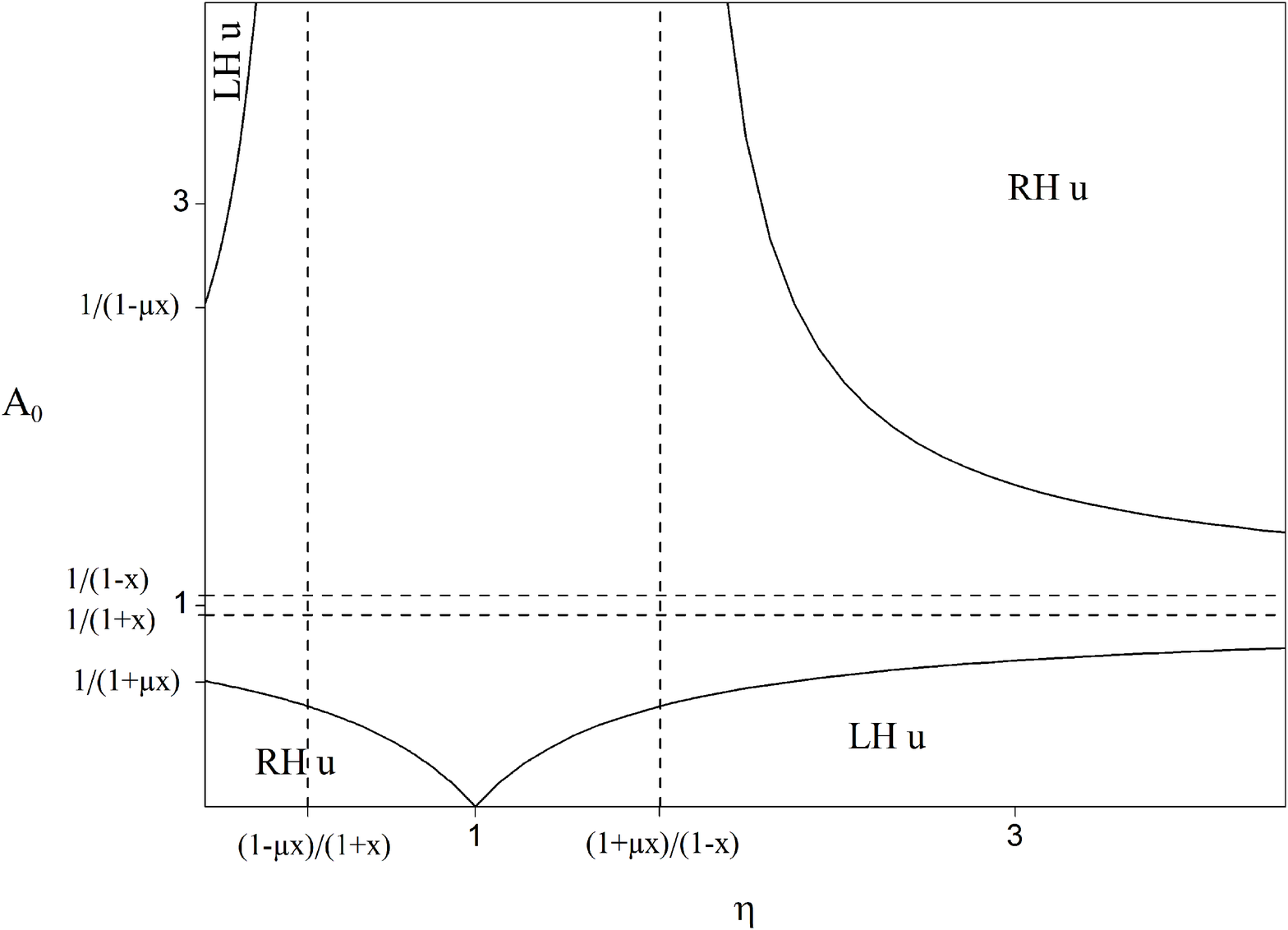}
\end{center}
\caption{Anisotropy diagram for LH polarized Alfven and proton
  cyclotron waves and for RH polarized Alfven and low phase speed
  Whistler waves for the case of equal electron and proton temperature
  anisotropies $A_e=A_p=A_0$. $\eta =\bp /(\w /\mu )$ denotes the
  normalized plasma beta. Unstable regions are marked by "u",}
\end{figure}
We remark at this point that the threshold condition for the LH
Alfven-proton cyclotron branch shown in Fig. 2 does not agree with the
AIC threshold condition plotted in Fig. 1 of \citet{Bale2009}.
\subsection{Right-handed polarized Alfven waves and low phase speed Whistler waves}
In the range of the RH polarized Alfven and Whistler waves below $x\ll
\mi2 $ the function $f_+(x)\simeq -1$ is negative and practically
constant, so that the instability condition (\ref{e9}) together with
$W_+=-\w /2\bp $ becomes

\bdm
\left({1\over A_0}-1\right)\left[(\m32 +1)\tanh \left[{\w \over 2\bp }\right]-(\m32 -1)\right]
\ebe
>x\mu \left[(\mpm +1)-(\mpm -1)\tanh \left[{\w \over 2\bp }\right]\right]
\label{e26}
\ee
With the approximation (\ref{e16}) we derive for condition (\ref{e26})

\be
({1\over A_0}-1)(1-\eta )>x(\mu +\eta )
\label{e27}
\ee
For $\bp <\bc $ the instability condition (\ref{e27}) becomes

\be
A_0(\bp <\bc )<{1-\eta \over (1+\mu x)-(1-x)\eta }
\label{e28}
\ee
Likewise, for $\bc <\bp <43\bc $ we obtain 

\be
A_0(\bp >\bc )>{\eta -1\over (1-x)\eta -(1+\mu x)}
\label{e29}
\ee
For plasma beta values equal to the critical value ($\eta =1$), the
instability condition (\ref{e19}) cannot be fulfilled.  The two
instability conditions (\ref{e28}) and (\ref{e29}) for the RH
polarized Alfven and low phase speed Whistler waves are illustrated in
Fig. 2.
\subsection{Right-handed polarized high phase speed Whistler waves and electron cyclotron waves}
In the range of high speed Whistler and electron cyclotron waves $x\gg
\mi2 $ the function $f_+(x)\simeq \mu x^2$ is positive.  The
instability condition (\ref{e9}) together with $W_+=\w f_+(x)/(2\bp
)=\mu \w x^2/(2\bp )$ then becomes

\bdm
\left(1-{1\over A_0}\right)\left[(\m32 -1)+(\m32 +1)\tanh \left[{\mu \w x^2\over 2\bp }\right]\right]>
\edm
\be
x\mu \left[(\mpm +1)+(\mpm -1)\tanh \left[{\mu \w x^2\over 2\bp }\right]\right]
\label{e30}
\ee
With the approximation (\ref{e16}) we derive for condition (\ref{e30})

\be
\left(1-{1\over A_0}\right)\left[1-{\bp \over \mu \w x^2}\right]>x\left[1+{\bp \over \mu \w x^2}\right].
\label{e31}
\ee
Because $x\gg \mi 2$ we find that 

\be
{\bp \over \mu \w x^2}\ll {\bp \over \w }\ll \mu ^{-1}
\label{e32},
\ee
so that 

\be
1-{1\over A_0}>x{1+{\bp \over \mu \w x^2}\over 1-{\bp \over \mu \w x^2}}\simeq x[1+{2\bp \over \mu \w x^2}]
\label{e33},
\ee
leading to 

\be
A_0>{1\over 1-x[1+{2\bp \over \mu \w x^2}]}
\label{e34}
\ee
We now discuss the instability conditions for individual wave modes.
\section{RH and LH polarized Alfven waves}
For $w\gg \mu ^{1/2}> 21.4$ we insert the Alfven wave dispersion
relation (\ref{d9}) in the conditions (\ref{e23}), (\ref{e25}),
(\ref{e28}) and (\ref{e29}). For $\bp <\bc $, corresponding to $\eta
<1$, we obtain for the RH polarized Alfven waves

\be
A_{0,RH}(\bp <\bc )<{w(1-\eta )\over (w+\mpm )-(w-\mi2 )\eta }={w(\bc -\bp )\over (w+\mpm )\bc -(w-\mi2 )\bp }
\label{f1}
\ee
and the LH polarized Alfven waves

\be
A_{0,LH}(\bp <\bc )>{w(1-\eta )\over (w-\mpm )-(w+\mi2 )\eta }={w(\bc -\bp )\over (w-\mpm )\bc -(w+\mi2 )\bp }
\label{f2}
\ee
Likewise, for $\bp >\bc $, corresponding to $\eta >1$, we obtain for
the RH polarized Alfven waves

\be
A_{0,RH}(\bp >\bc )>{w(\eta -1)\over (w-\mpm )\eta -(w+\mi2 )}={w(\bp -\bc )\over (w-\mpm )\bp -(w+\mi2 )\bc }
\label{f3}
\ee
and the LH polarized Alfven waves

\be
A_{0,LH}(\bp >\bc )<{w(\eta -1)\over (w+\mi2 )\eta -(w-\mpm )}={w(\bp -\bc )\over (w+\mi2 )\bp -(w-\mpm )\bc } 
\label{f4}
\ee
These four Alfvenic instability conditions are shown in Fig. 3. Note,
as indicated in the figure, that the limiting anisotropy values and
parallel plasma betas for $\bp \to 0$ and $\bp \to \infty$ and $A_0\to
\infty$, respectively, are solely determined by the mass ratio $\mu $
and the wavenumber-dependent plasma frequency phase speed $w=\wpe
/kc$.

\begin{figure}[t]
\begin{center}
\includegraphics[width=160mm]{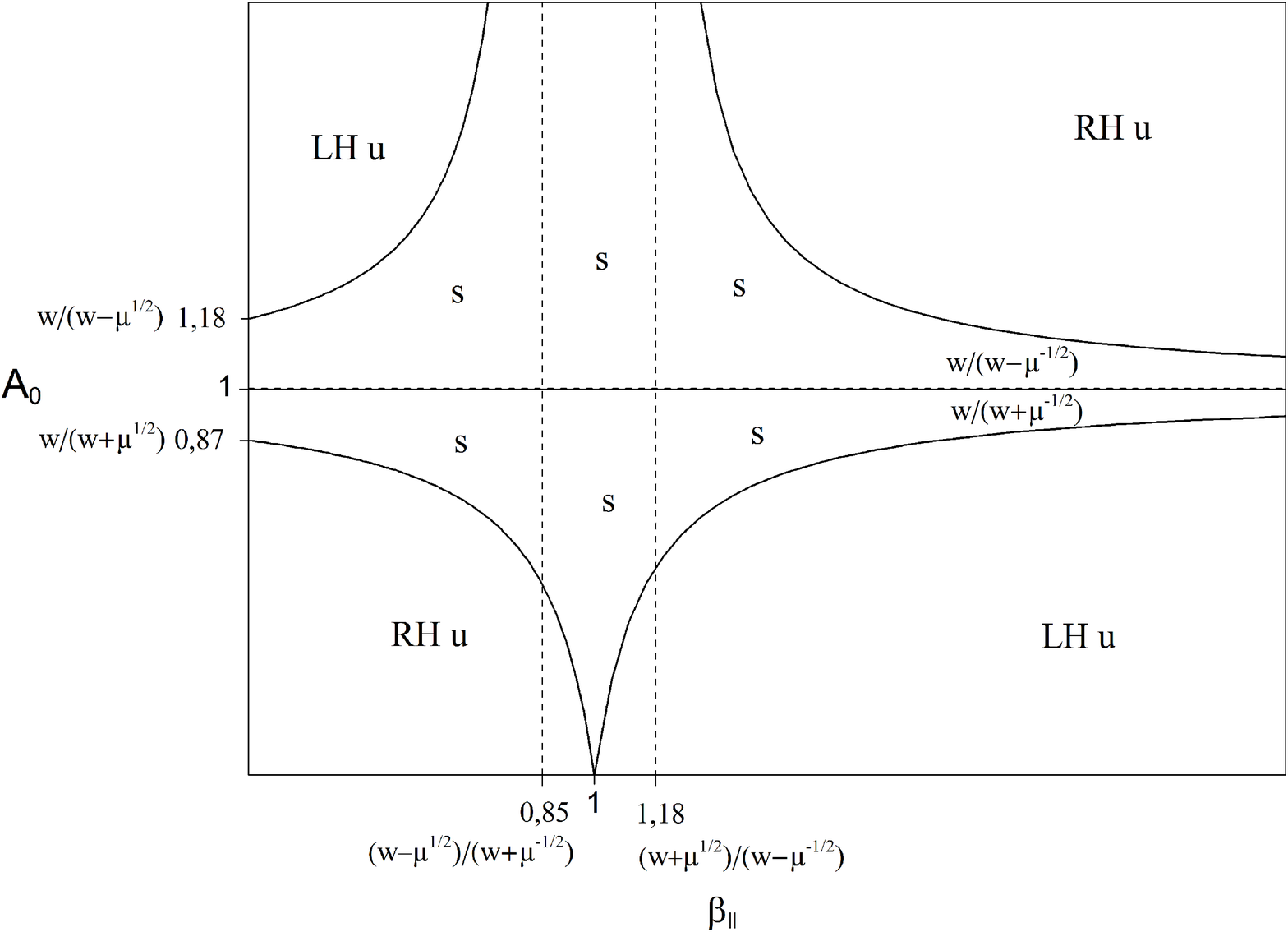}
\end{center}
\caption{Anisotropy diagram for LH and RH polarized Alfven waves for
  the case of equal electron and proton temperature anisotropies
  $A_e=A_p=A_0$. Stable regions are marked by "S", unstable regions
  are marked by "u". A value of $w=280$ is adopted so that the results
  hold for parallel plasma beta values $\bp <5.4$.}
\end{figure}

The properties of the weakly amplified polarized LH and RH polarized
Alfven mode are summarized in Table 1 and 2, respectively.
 
\begin{table}
\centering
\begin{tabular}{|l|l|}
\hline
Real phase speed range & $R\ll b/\mu =w\Tn /(\mu \bp ^{1/2})$ \\
\hline
Parallel plasma beta range &  $\Tn ^2\ll \bp \ll \w /\mu $ \\
\hline
Dispersion relation & $x=R/b={1\over \mu ^{1/2}w}$ \\
\hline
Existence condition  & $w\gg \mu ^{1/2}=43>\sqrt{(\mu /4)-1}=21.4$ \\
\hline
Instability conditions: & $A_0\notin \left[{w\over w+\mi2 },{w\over w-\mpm }\right]=\left[{43w\over 1+43w},{w\over w-43}\right]$ \\
\hline
for $\bp <\bc={\w \over (1+\m32 )} $ & $A_0>{w(\bc -\bp )\over (w-\mpm )\bc -(w+\mi2 )\bp } $ \\
\hline
for $\bp >\bc $ & $A_0<{w(\bp -\bc )\over (w+\mi2 )\bp -(w-\mpm )\bc }$ \\
\hline
\end{tabular}
\caption{Properties of weakly amplified LH polarized Alfven wave mode}
\end{table}

\begin{table}
\centering
\begin{tabular}{|l|l|}
\hline
Real phase speed range & $R\ll b/\mu =w\Tn /(\mu \bp ^{1/2})$ \\
\hline
Parallel plasma beta range &  $\Tn ^2\ll \bp \ll \w /\mu $ \\
\hline
Dispersion relation & $x=R/b={1\over \mu ^{1/2}w}$ \\
\hline
Existence condition  & $w\gg \mu ^{1/2}=43>\sqrt{(\mu /4)-1}=21.4$ \\
\hline
Instability conditions: & $A_0\notin \left[{w\over w+\mpm },{w\over w-\mi2 }\right]=\left[{w\over w+43},{43w\over 43w-1}\right]$ \\
\hline
for $\bp <\bc={\w \over (1+\m32 )} $ & $A_0<{w(\bc -\bp )\over (w+\mpm )\bc -(w-\mi2 )\bp } $  \\
\hline
for $\bp >\bc $ & $A_0>{w(\bp -\bc )\over (w-\mpm )\bp -(w+\mi2 )\bc }$ \\
\hline
\end{tabular}
\caption{Properties of weakly amplified RH polarized Alfven wave mode}
\end{table}

\section{Proton cyclotron waves}
For $w\ll 21.4$ we insert the dispersion relation (\ref{d6}) for
proton cyclotron waves in the instability conditions (\ref{e23}) and
(\ref{e25}) yielding

\be
A_{0}(\bp <\bc )>{\mu (1-\eta )\over 1+\w -(\mu +1)\eta }={\mu (\bc -\bp )\over (1+\w )\bc -(\mu +1)\bp }
\label{f5}
\ee
and

\be
A_{0}(\bp >\bc )<{\mu (\eta -1)\over (\mu +1)\eta -(1+\w )}={\mu (\bp -\bc )\over (\mu +1)\bp -(1+\w )\bc } 
\label{f6}
\ee
These instability conditions cannot be fulfilled for anisotropies in
the interval $A_0\in [\mu /(1+\mu ), \mu /(1+\w )]$, including the
isotropic case $A_0=1$. The proton cyclotron instability conditions
are illustrated in Fig. 4 as a function of the parallel plasma beta
$\bp $.

\begin{figure}[t]
\begin{center}
\includegraphics[width=160mm]{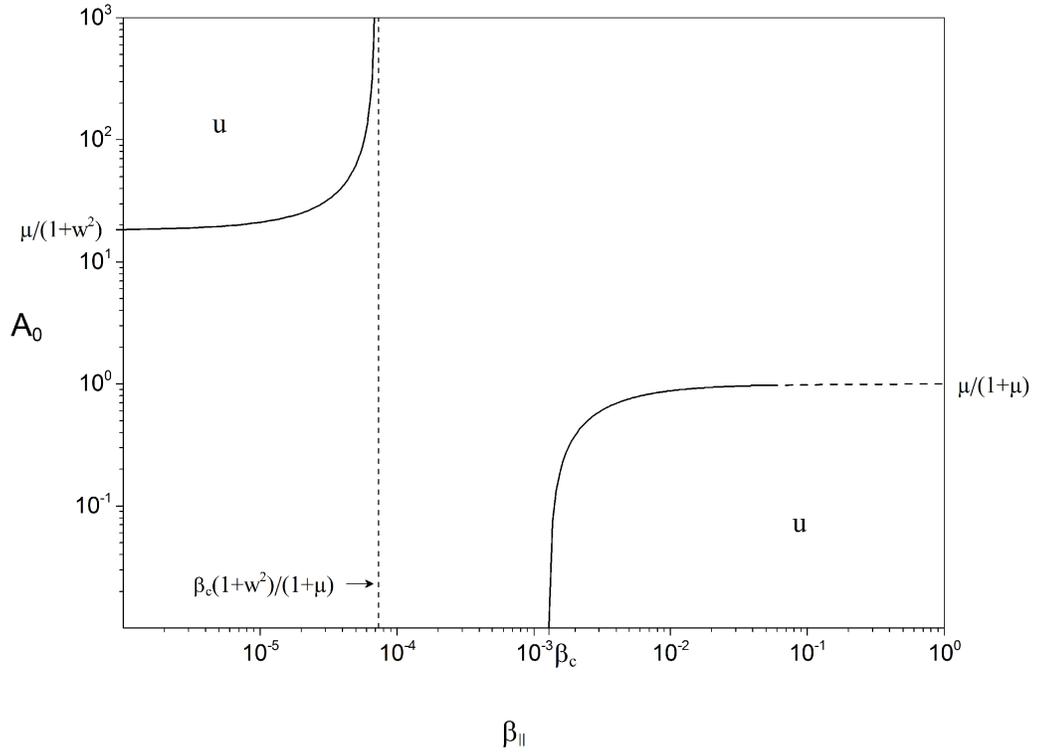}
\end{center}
\caption{Anisotropy diagram for LH polarized proton cyclotron waves
  for the case of equal electron and proton temperature anisotropies
  $A_e=A_p=A_0$. Unstable regions are marked by "u", A value of $w=10$
  is adopted so that the results hold for parallel plasma beta values
  $\bp <0.054$.}
\end{figure}

\begin{table}
\centering
\begin{tabular}{|l|l|}
\hline
Real phase speed range & $R\ll b$  \\
\hline
Parallel plasma beta range & $\Tn ^2\ll \bp \ll {\w \over \mu } $ \\
\hline
Dispersion relation & $x=R/b={1\over \mu }[1-{1+\w \over \mu }]$ \\
\hline
Existence condition  & $w\ll \sqrt{(\mu /4)-1}=21.4$ \\
\hline
Instability conditions: & $A_0\notin \left[{\mu \over 1+\mu }, {\mu \over 1+\w }\right]$ \\
\hline
for $\bp <\bc={\w \over (1+\m32 )} $ &   $A_0>{\mu (\bc -\bp )\over (1+\w )\bc -(\mu +1)\bp }$ \\
\hline
for $\bp >\bc $ & $A_0<{\mu (\bp -\bc )\over (\mu +1)\bp -(1+\w )\bc }$ \\
\hline
\end{tabular}
\caption{Properties of weakly amplified LH polarized proton cyclotron
  mode}
\end{table}

The properties of the weakly amplified proton cyclotron mode are
summarized in Table 3.

\section{Right-handed polarized low phase speed Whistler waves}
The Whistler-electron cyclotron dispersion relation (\ref{d7}) yields
values of $x\ll \mi2 $ provided

\be
{1\over 1+\w }\ll \mi2 ,
\label{f7}
\ee
which is equivalent to 

\be
(\mpm -1)^{1/2}=6.5\ll w\ll 21.4
\label{f8}
\ee
The instability conditions (\ref{e28}) and (\ref{e29}) then reduce to 

\be
A_0(\bp <\bc )<{(1+\w )(1-\eta )\over (1+\mu +\w )-\w \eta }={(1+\w )(\bc -\bp )\over (1+\mu +\w )\bc -\w \bp }
\label{f9}
\ee
and 

\be
A_0(\bp >\bc )>{(1+\w )(\eta -1)\over \w \eta -(1+\mu +\w )}={(1+\w )(\bp -\bc )\over \w \bp -(1+\mu +\w )\bc }
\label{f10}
\ee
These instability condition cannot be fulfilled for anisotropies in
the interval

\be
{1+\w \over 1+\w +\mu }\le A_0\le {1+\w \over \w },
\label{f11}
\ee
including the isotropic case $A_0=1$. The RH low phase speed Whistler
wave instability conditions are illustrated in Fig. 5 as a function of
the parallel plasma beta $\bp $.

\begin{figure}[t]
\begin{center}
\includegraphics[width=160mm]{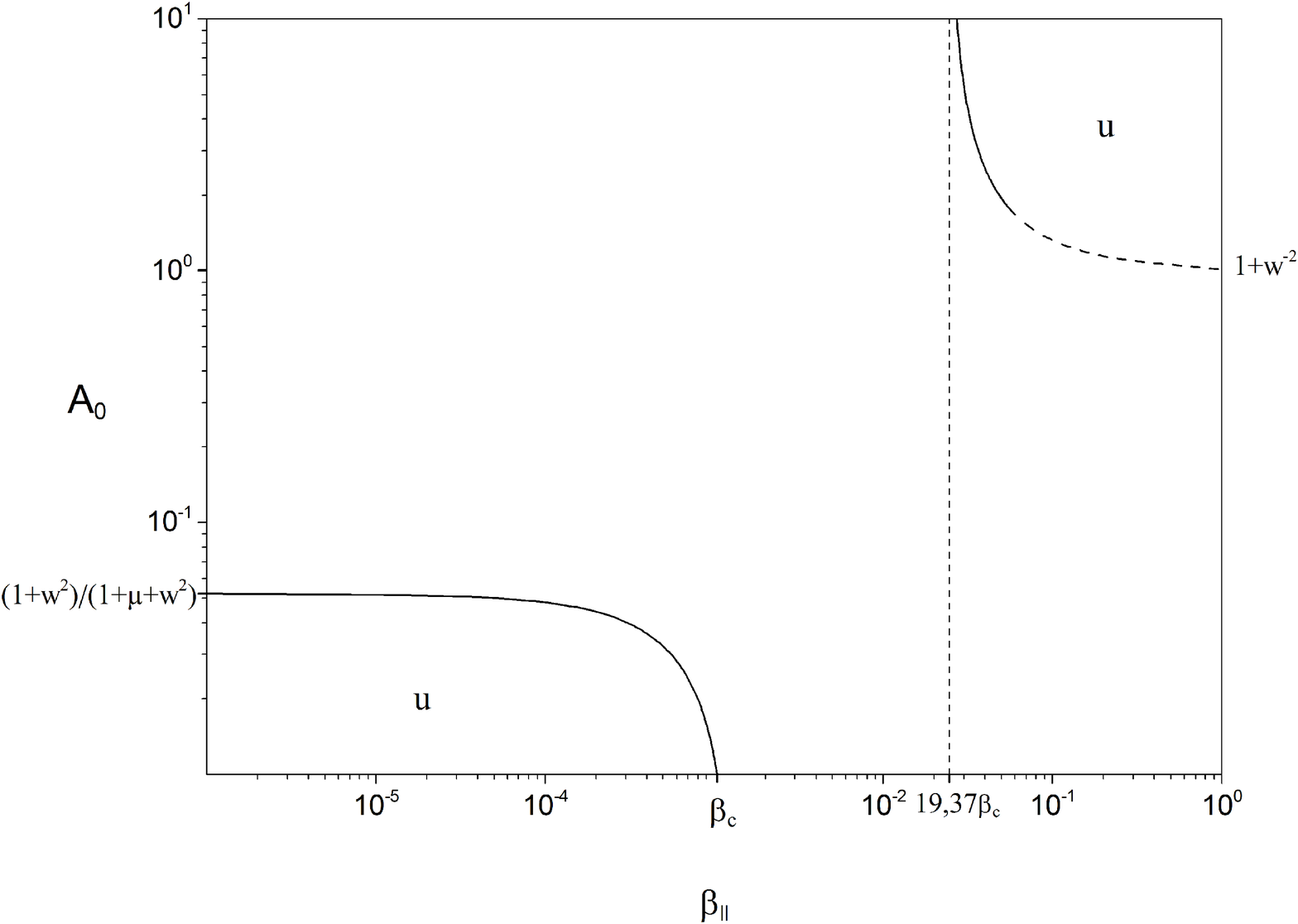}
\end{center}
\caption{Anisotropy diagram for RH polarized low phase speed Whistler
  waves for the case of equal electron and proton temperature
  anisotropies $A_e=A_p=A_0$. Unstable regions are marked by "u". A
  value of $w=10$ is adopted so that the results hold for parallel
  plasma beta values $\bp <0.054$.}
\end{figure}
The properties of the weakly amplified RH polarized low phase speed
Whistler mode are summarized in Table 4.

\begin{table}
\centering
\begin{tabular}{|l|l|}
\hline
Real phase speed range & ${b\over \mu }\ll R\ll b\mi2 $  \\
\hline
Parallel plasma beta range & $\Tn ^2\ll \bp \ll \w /\mu  $ \\
\hline
Dispersion relation & $x=R/b={1\over 1+\w }$ \\
\hline
Existence condition  & $6.5\ll w\ll 21.4$ \\
\hline
Instability conditions: & $A_0\notin \left[{1+\w \over 1+\mu +\w },{1+\w \over \w }\right]$ \\
\hline
for $\bp <\bc={\w \over (1+\m32 )} $ & $A_0<{(1+\w )(\bc -\bp )\over (1+\mu +\w )\bc -\w \bp }$   \\
\hline
for $\bp >\bc $ &  $A_0>{(1+\w )(\bp -\bc )\over \w \bp -(1+\mu +\w )\bc }$ \\
\hline
\end{tabular}
\caption{Properties of weakly amplified RH polarized low phase speed Whistler mode}
\end{table}

\section{Right-handed polarized electron cyclotron waves and high phase speed Whistler waves}
The Whistler-electron cyclotron dispersion relation (\ref{d7}) yields
values of $x\gg \mi2 $ provided

\be
{1\over 1+\w }\gg \mi2 ,
\label{f12}
\ee
which is equivalent to 

\be
w\ll (\mpm -1)^{1/2}=6.5
\label{f13}
\ee
Inserting the Whistler-electron cyclotron dispersion relation
(\ref{d7}) yields for the instability condition (\ref{e34})

\be
A_0>{\mu \w (1+\w )\over \mu w^4-2(1+\w )^2\bp },
\label{f14}
\ee
which is illustrated in Fig. 6.

\begin{figure}[t]
\begin{center}
\includegraphics[width=160mm]{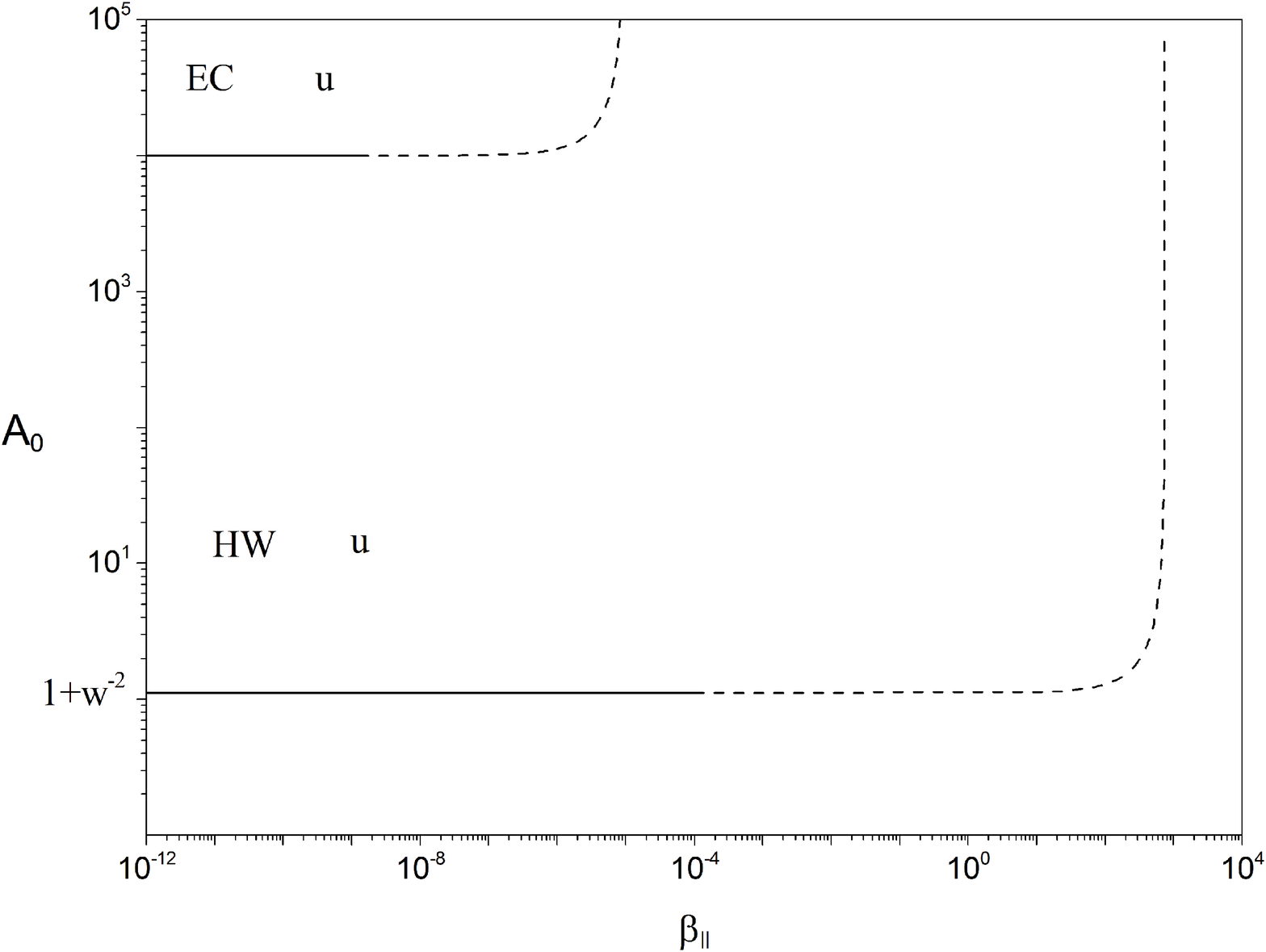}
\end{center}
\caption{Anisotropy diagram for RH polarized high phase speed Whistler
  and electron cyclotron waves for the case of equal electron and
  proton temperature anisotropies $A_e=A_p=A_0$. Unstable regions are
  marked by "u", Values of $w=3$ and $w=0.01$ are adopted for high
  phase speed Whistlers and electron cyclotron waves, respectively, so
  that the results hold for parallel plasma beta values below $5\cdot
  10^{-3}$ and $5\cdot 10^{-6}$, respectively.}
\end{figure}
The properties of the weakly amplified RH polarized high phase speed
Whistler and electron cyclotron modes are summarized in Table 5.

\begin{table}
\centering
\begin{tabular}{|l|l|}
\hline
Real phase speed range & $b\mi2 \ll R\ll b$  \\
\hline
Parallel plasma beta range & $\Tn ^2\ll \bp \ll \w /\mu $ \\
\hline
Dispersion relation & $x=R/b={1\over 1+\w }$ \\
\hline
Existence condition  & $w\ll 6.5$ \\
\hline
Instability condition: & \\
\hline
for $\bp < \w /\mu $ & $A_0>{\mu \w (1+\w )\over \mu w^4-2(1+\w )^2\bp } $   \\
\hline
\end{tabular}
\caption{Properties of weakly amplified RH polarized high phase speed
  Whistler and electron cyclotron wave mode}
\end{table}

\section{Application to solar wind fluctuations}
The solar wind magnetic fluctuations measured by \citet{Bale2009}
near 1 AU have wavenumbers

\be
k\simeq \a /\rho _p
\label{s1}
\ee
with $\a =0.56\pm 0.32$ and the thermal proton gyroradius $\rho
_p=4.23\cdot 10^6T_5^{1/2}B_4^{-1}$ cm, where we adopt an
interplanetary magnetic field value $B=10^{-4}B_4$ Gauss and a
temperature $T=10^5T_5$ K. With the solar wind particle density
$n_e=10^2n_2$ cm$^{-3}$ we find that the plasma frequency phase speed
(\ref{a8})

\be
w={79.5\over \a }{(T_5n_2)^{1/2}\over B_4}={142\over 1\pm 0.57}{(T_5n_2)^{1/2}\over B_4}
\label{s2}
\ee
covers the interval 

\be
90{(T_5n_2)^{1/2}\over B_4}\le w\le 330{(T_5n_2)^{1/2}\over B_4}
\label{s3}
\ee
The electron gyrofrequency phase speed (\ref{a81})

\be
b=3.13\cdot 10^{-3}w{B_4\over n_2^{1/2}}
\label{s4}
\ee
is indeed much smaller than the electron plasma frequency phase speed
$w$ justifying the high density plasma approximation $w\gg b$.

According to the existence conditions of the different instabilities
for low plasma beta values summarized in Tables 1-5, only the
left-handed and right-handed polarized Alfven wave instabilities can
operate in the $w$-interval (\ref{s3}). Fig. 3 therefore applies to
this application.  Fig. 3 was calculated for a value of $w=280$ lying
within the range (\ref{s3}) of the observed solar wind
turbulence. $w=280$ yields for the critical plasma beta $\bc =\w
/(1+\mu ^{3/2})=0.997$.

Besides the mass ratio $\mu $ the Alfvenic instability threshold
diagram shown in Fig. 3 is controlled by the single observed plasma
parameter $w$. These two quantities define the characteristic plasma
beta $\bc =\w /\mu ^{3/2}$ and the parameter

\be
a={\mpm -\mi 2\over w-\mi2 }\simeq {\mpm \over w}
\label{s5}
\ee
As illustrated in Fig. 7 all asymptotic and characteristic values of
the threshold anisotropies and the characteristic plasma beta values
in the case of LH and RH polarized Alfven waves are solely determined
by $a$ and $\bc $.

\begin{figure}[t]
\begin{center}
\includegraphics[width=160mm]{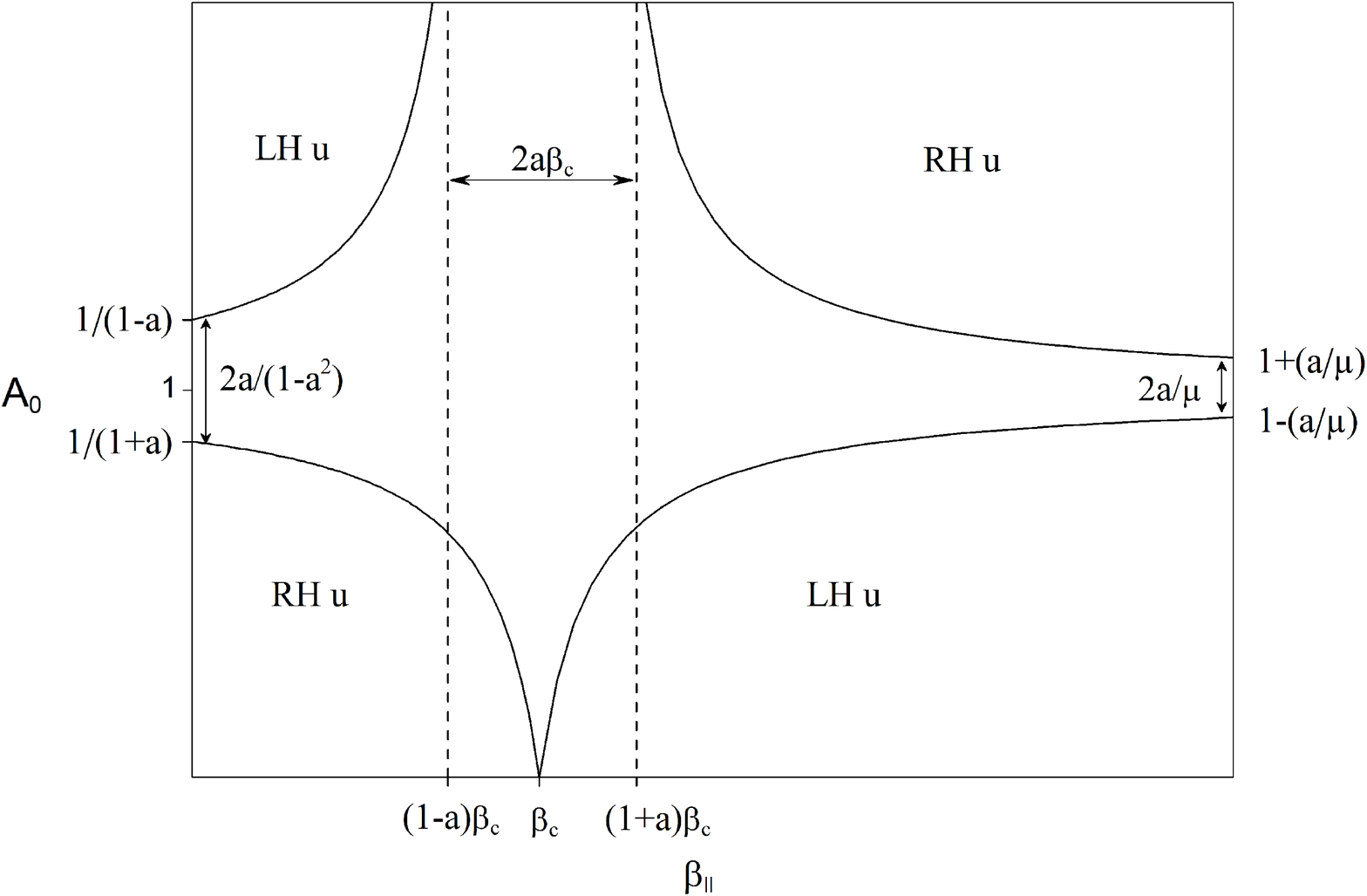}
\end{center}
\caption{Anisotropy diagram for LH and RH polarized Alfven waves in
  terms of the plasma parameters $\bc $ and $a=\mu ^{1/2}/w$.
  Unstable regions are matrked by "u",}
\end{figure}

\section {Summary and conclusions}
We rigorously studied the dispersion relations of weakly amplified
fluctuations with wave vectors $\vec{k}\times \vec{B}_0=0$ in an
anisotropic bi-Maxwellian magnetized proton-electron plasma by the
appropriate Taylor expansion of the plasma dispersion function.
Apparently for the first time, for equal parallel electron and proton
temperatures, a general analytical instability condition (\ref{e8}) is
derived that holds for different values of the electron ($A_e$) and
proton ($A_p$) temperature anisotropies. We determine the conditions
for which the weakly amplified LH-handed polarized
Alfven-proton-cyclotron and RH-handed polarized
Alfven-Whistler-electron-cyclotron branches can be excited.  For
different regimes of the electron plasma frequeny phase speed $w=\wpe
/(kc)$ these branches reduce to the RH and LH polarized Alfven waves,
RH polarized high- and low-phase speed Whistler, RH polarized proton
and LH polarized electron cyclotron modes. The properties of the
individual modes are summarized in Tables 1-5. Analytic instability
threshold conditions are derived in terms of the combined temperature
anisotropy $A=T_{\perp }/T_{\parallel }$, the parallel plasma beta
$\bp =8\pi n_ek_BT_{\parallel }/B^2$ and the electron plasma frequency
phase speed $w=\wpe /(kc)$ for each mode.
 
For large values of $w\gg 21.4$, corresponding to small wavenumbers
$k\ll \wpe /21.4c$, RH and LH polarized Alfven waves are excited in
different regions of the temperature anisotropy (A)-versus parallel
plasma beta ($\bp $)-parameter plane.  For appropriate temperature
anisotropies, RH polarized low and high phase speed Whistler waves are
excited for intermediate values of $6.5\ll w\ll 21.4$ and $1\ll w\ll
6.5$, respectively. At values $w\ll 21.4$ LH polarized proton
cyclotron waves can be excited, whereas RH polarized electron
cyclotron waves can be excited at small values of $w\le 1$. In
agreement with the general theorem of \citet{br90} on the electromagnetic
stability of isotropic plasma populations none of these modes can be
excited for isotropic plasma distributions ($A_e=A_p=1$).

We apply the results of our instability study to the observed solar
wind magnetic turbulence \citep{Bale2009}, corresponding to values
of $90\le w\le 330$. According to the existence conditions of the
different instabilities, only the left-handed and right-handed
polarized Alfven wave instabilities can operate here. Besides the
electron-proton mass ratio $\mu =1836$, the Alfvenic instability
threshold conditions are controlled by the single observed plasma
parameter $w$. All asymptotic and characteristic values of the
threshold anisotropies and the characteristic plasma beta values are
solely determined by the values of $a=43/w$ and $\bc =w^2/\mu
^{3/2}$. Comparing the Alfvenic instability diagram shown in Fig. 7
with the observations shown in Fig. 1 of \citet{Bale2009}, one
notices that the main characteristic properties of the observed solar
wind fluctuations are well reproduced by the instability conditions of
parallel propagating Alfven waves. Especially, the observed
confinement limits at small parallel plasma beta values are
explained. It remains to be investigated in future work how \wp and/or
obliquely propagating weakly amplified modes add to this diagram.
\acknowledgments We thank the referee for his constructive comments
that helped to improve the manuscript. This work was supported by the
Deutsche Forschungsgemeinschaft through grants Schl 201/19-1 and Schl
201/21-1.

\bibliography{sts_bib}

\end{document}